# The MC-QTAIM analysis reveals an exotic bond in the coherently quantum superposed Malonaldehyde

Mohammad Goli[1] and Shant Shahbazian[2,*]

[1]School of Nano Science, Institute for Research in Fundamental Sciences (IPM), Tehran 19395-5531, Iran,

Email: m_goli@ipm.ir

[2]Department of Physics, Shahid Beheshti University, Evin, Tehran, Iran,

E-mail: sh_shahbazian@sbu.ac.ir

**Abstract**

The proton between the two oxygen atoms of the malonaldehyde molecule experiences an effective double-well potential in which the proton's wavefunction is delocalized between the two wells. Herein we employed the state-of-the-art multi-component quantum theory of atoms in molecules partitioning scheme to obtain the molecular structure, i.e. atoms in molecules and bonding network, from the superposed ab initio wavefunctions of malonaldehyde. In contrast to the familiar clamped-proton portrayal of malonaldehyde, in which the proton forms a hydrogen basin, for the superposed states the hydrogen basin disappears and two novel hybrid oxygen-hydrogen basins appear instead, with an even distribution of the proton population between the two basins. The interaction between the hybrid basins is stabilizing thanks to an unprecedented mechanism. This involves the stabilizing classical Coulomb interaction of the one-proton density in one of the basins with the one-electron density in the other basin. This stabilizing mechanism yields a bond foreign to the known bonding modes in chemistry.

## I. Introduction

The superposition principle is probably the most important principle of quantum mechanics with many theoretical ramifications like the exotic quantum correlations and the entanglement, which are quite intriguing from the classical physicist viewpoint.[1,2] In quantum chemistry and the electronic structure theory the superposition principle manifests itself by various linear combinations of atomic orbitals,[3] molecular orbitals,[4] and Slater determinants.[5,6] However, the majority of these manifestations include only the electronic states which are derived from the electronic Hamiltonian that treats electrons as quantum particles and the nuclei as clamped point charges. Although it is generally highlighted that the electronic Hamiltonian is just an approximation based on the adiabatic separation of the electronic and nuclear motions,[7] its almost-universal computational usage conceals its basic shortcoming, namely the total dismissal of the superposition of the nuclear states.[8,9]

To consider the superposed nuclear states, one may solve the nuclear Schrödinger equation using the potential energy surface (PES) derived from the solution of the electronic Schrödinger equation.[10] However, this approach has rarely been used to challenge the viewpoint of the concept of "molecular structure" in which electrons are delocalized throughout the molecule while nuclei are conceived as point charges vibrating locally around their equilibrium positions.[11–14] Some simple examples that may challenge this prototypical viewpoint are the inversion motion of pyramidal molecules, e.g. ammonia, and the inter-conversion of chiral isomers where both are usually modeled by an effective double-well potential.[15,16] In these cases, there are two molecular structures each represented by a well and separated by a finite potential barrier; however, the true ground state wavefunction is indeed a superposition of the wavefunctions of these two structures.

[17–19] Evidently, in the superposed ground state certain, if not all, nuclei are "delocalized" in two (or sometimes more) spatial positions, and the simple picture of a "localized" nucleus excursing around an equilibrium spatial point is no longer applicable.

The question arises why such superposed states are not usually observable in chemical experiments. Various physical mechanisms, e.g. the environmentally induced decoherence,[20–23] and certain mathematical constraints, e.g. the spontaneous symmetry breaking or the superselection rules,[24–27] have been utilized to explain the subtlety of observation of such states. Nonetheless, currently, there is no doubt that it is experimentally feasible to construct such superposed states in a controllable way in sufficiently isolated situations.[28,29]

One interesting example of such systems is the malonaldehyde molecule wherein the proton between the two oxygen atoms experiences an effective double-well potential.[30,31] The two indistinguishable conformers, depicted in Figure 1, are local minima on the PES, with **Cs** geometrical symmetry, that are derivable from the electronic Schrödinger equation.[32] The proton transfer reaction involves a transition state, depicted in Figure 1, with **C2v** geometrical symmetry.[32–34]

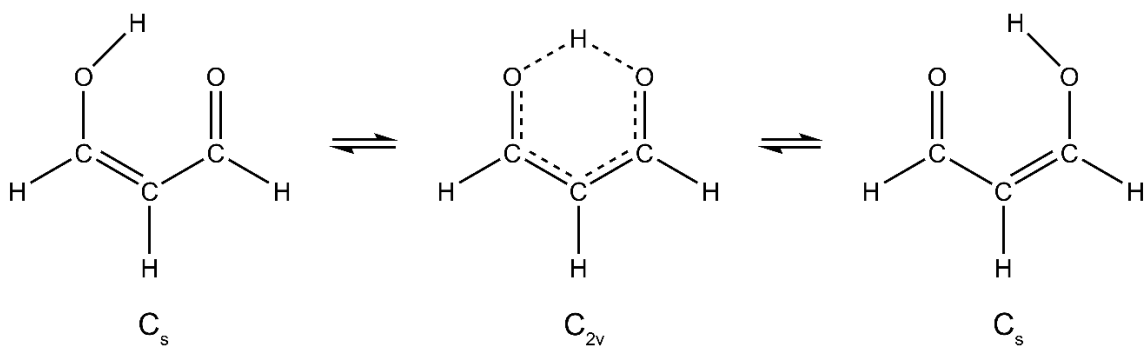

**Figure 1** Schematic representation of the intra-molecular hydrogen transfer in malonaldehyde between the two conformers, **Cs**, passing through the transition state, **C2v**.

The proton transfer between the two local minima may happen through the tunneling process, which has been extensively studied both theoretically,[35–62] and experimentally, [63–83] in the last decades. The tunneling mechanism implies that if **Cs** structures are prepared as stationary states, an oscillatory time-dependent non-stationary process must transform one conformer to the other. [30,31] The alternative "delocalized" scenario emerges if the proton is treated as a quantum particle from the outset in the ab initio calculations. The resulting two-component Hamiltonian, which contains the kinetic energy operators of the proton and electrons simultaneously, belongs to the realm of the multi-component (MC) quantum chemistry.[84–90] In that case, the resulting two-component (TC) wavefunctions of the ground and first excited states contain the electronic and protonic variables concurrently. If the MC Schrödinger equation is solved with sufficient accuracy,[91–94] the probability densities of the proton's wavefunctions in both states reveal two peaks at two distinct positions between the two oxygen atoms. These superposed structures are depicted schematically as **4**, the ground state, and **5**, the first excited state, in Figure 2 and compared to structures **1** to **3**. In the latter structures, the quantum proton is automatically confined to a spatial excursion around a reference point at the minima or the maximum on the effective PES depicted in Figure 2 and, these structures are the TC analogs of **Cs** and **C2v**.

The molecular structures at their root are the networks of atoms in molecules (AIM) and their linking bonds. In line with this context, the quantum theory of atoms in molecules (QTAIM) is a quantum chemical partitioning methodology that divides a molecule exhaustively into atomic basins in the 3D real space and also partitions the molecular properties into the basin and inter-basin contributions.[95–97] Since the QTAIM partitioning algorithm solely employs the electronic wavefunctions as the inputs to derive AIM and their associated properties and interaction modes as the outputs, it fails to perform the AIM partitioning of the TC wavefunctions. To address this issue, the MC extension of the QTAIM, called MC-QTAIM, has recently been developed that is capable of partitioning the MC quantum systems by employing the ab initio MC wavefunctions.[98–117] Our goal in this study is to decipher the molecular structures of the superposed states of malonaldehyde, **4** and **5**, by applying the MC-QTAIM to the corresponding wavefunctions.

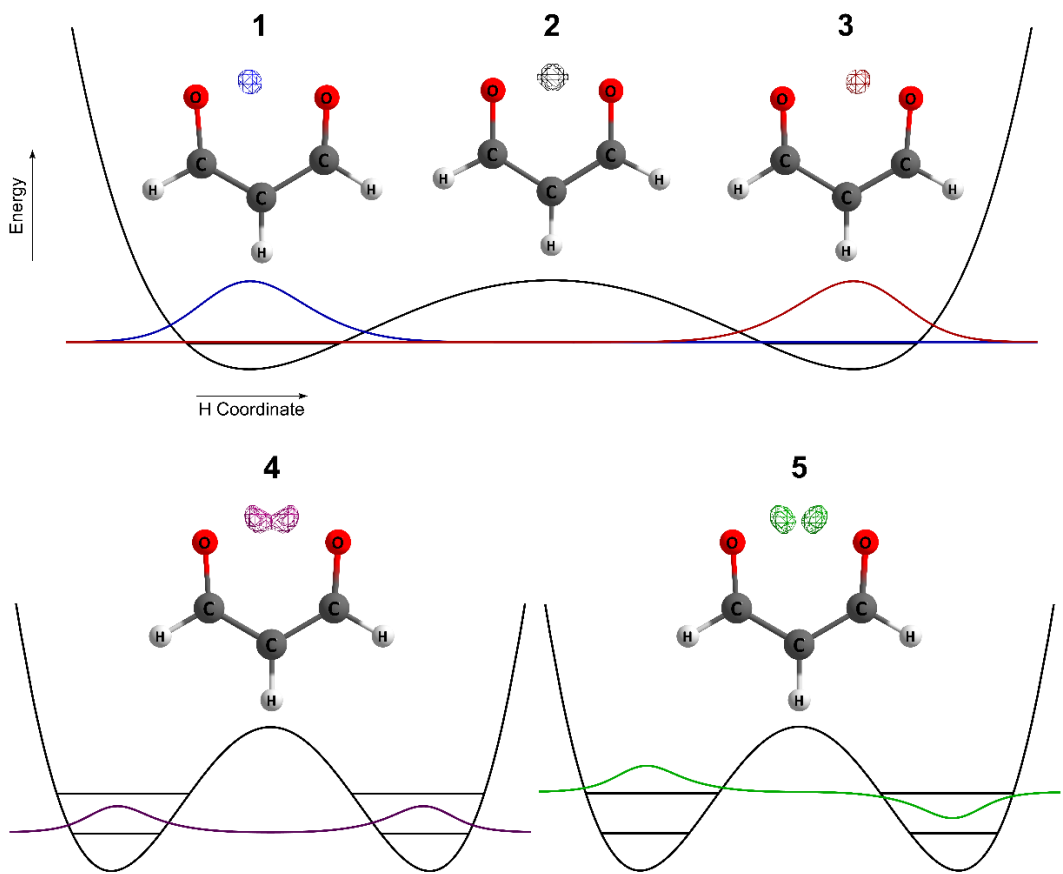

**Figure 2** Schematic depictions of malonaldehyde structures with the localized (**1**-**3**) and delocalized protons (**4** and **5**), and, the corresponding one-dimensional cross section of the effective potential energy surfaces experienced by the proton along the axis connecting the two oxygen nuclei, denoted also by the H Coordinate, and the one-proton wavefunctions (while a single color has been used to depict the wavefunction of **5**, the phase of wavefunction is different in the two wells). The effective potential is the potential energy that is used in the one-particle Schrödinger equation governing the proton to deduce proton's wavefunction in a computational procedure detailed in the computational details section.

## 2. Theory and computational details

### 2.1. The MC-QTAIM

In the last decade, the QTAIM scheme has been extended to the partitioning of the MC quantum systems, and the input to the extended algorithm, called MC-QTAIM,[98–117] are ab initio MC wavefunctions derived from solving the MC Schrödinger equation (written in atomic units):

$$\hat{H}_{MC}\Psi_{MC}\left(\left\{\vec{x}_{1,1},...,\vec{x}_{1,N_1}\right\},...,\left\{\vec{x}_{s,1},...,\vec{x}_{s,N_s}\right\};\left\{\vec{R}_\alpha\right\}\right)$$

$$= E_{MC}\left(\left\{\vec{R}_\alpha\right\}\right)\Psi_{MC}\left(\left\{\vec{x}_{1,1},...,\vec{x}_{1,N_1}\right\},...,\left\{\vec{x}_{s,1},...,\vec{x}_{s,N_s}\right\};\left\{\vec{R}_\alpha\right\}\right),$$

$$\hat{H}_{MC} = \sum_n^s (-1/2m_n)\sum_i^{N_n}\nabla^2_{n,i} + \sum_n^s\sum_i^{N_n}\sum_{j>i}^{N_n}\frac{q_n^2}{|\vec{r}_{n,i}-\vec{r}_{n,j}|} + \sum_n^s\sum_{m>n}^s\sum_i^{N_n}\sum_j^{N_m}\frac{q_n q_m}{|\vec{r}_{n,i}-\vec{r}_{m,j}|} + \sum_n^s\sum_i^{N_n}\sum_\alpha^Q\frac{Z_\alpha q_n}{|\vec{R}_\alpha-\vec{r}_{n,i}|}$$

(1)

This equation governs a system with $s$ types of distinguishable quantum particles, with spin-spatial variables $\vec{x}_{n,i} = (\vec{r}_{n,i}, \sigma_{n,i})$ for the *i*-th particle of the $n-th$ type, where there are $N_n$ number of particles with charge, $q_n$, and mass, $m_n$, interacting via the Coulomb law with each other and the clamped nuclei; there are $Q$ number of the latter carrying $Z_\alpha$ charge and placed at $\vec{R}_\alpha$. The zero-flux equation of the MC-QTAIM partitioning algorithm, used to derive the boundaries of the atomic basins, is as follows:[105,117]

$$\vec{\nabla}\Gamma^{(s)}(\vec{r}).\vec{n}(\vec{r}) = 0, \quad \Gamma^{(s)}(\vec{r}) = \sum_n^s (m_1/m_n)\rho_n(\vec{r}) \quad (2)$$

Wherein, $\rho_n(\vec{r}) = N_n \int d\vec{x}_{1,1}...\int d\vec{x}_{n-1,N_n}\int d\sigma_{n,1}\int d\vec{x}_{n,2}....\int d\vec{x}_{s,N_s} \Psi^*_{MC}\Psi_{MC}$ is the one-particle density of $n-th$ type of particles.[105,117] The Gamma density, $\Gamma^{(s)}$, which replaces the one-electron density used within the context of the QTAIM, is a mass-scaled combined density that includes the contributions of all quantum particles in shaping the boundaries. The numbering of particle types relies on their mass and starts from the lightest to the heaviest particle type, e.g. for a system composed of electrons and protons as quantum particles, $\rho_1(\vec{r})$ and $\rho_2(\vec{r})$ are the electron and the proton one-densities, respectively. Some quantum particles, e.g. hydrogen isotopes, can form their own atomic basins, and the number of basins in an MC system is generally larger than the number of clamped nuclei, $P > Q$.[107,108] The one-particle properties, $\hat{M} = \sum_n^s \hat{M}_n = \sum_n^s \left( \sum_i^{N_n} \hat{m}_{n,i} \right)$, are partitioned into basin contributions as follows:[104,105]

$$\left\langle \hat{M} \right\rangle = \left\langle \sum_n^s \sum_i^{N_n} \hat{m}_{n,i} \right\rangle = \sum_k^P \tilde{M}(\Omega_k), \quad \tilde{M}(\Omega_k) = \sum_n^s M_n(\Omega_k) \tag{3}$$

Wherein, $M_n(\Omega_k) = N_n \int d\vec{x}_{1,1} ... \int d\vec{x}_{n-1,N_n} \int_{\Omega_k} d\vec{r}_{n,1} \int d\sigma_{n,1} \int d\vec{x}_{n,2} ... \int d\vec{x}_{s,N_s} \text{Re}\left[ \Psi_{MC}^* \hat{m}_{n,1} \Psi_{MC} \right]$

is the contribution of $n-th$ type of particles to the basin property $\hat{M}$. In principle, the basin contribution originates from all types of particles though in practice some heavy particles are localized around a certain position and merely contribute to the properties of the basins within which they are confined.[101,107–109,112] The partitioning of the two-particle properties, $\hat{G} = \sum_n^s \left( \sum_i^{N_n} \sum_{j>i}^{N_n} \hat{g}_{n,ij} \right) + \sum_n^s \sum_{m>n}^s \left( \sum_i^{N_n} \sum_j^{N_m} \hat{g}_{nm,ij} \right)$, is done as follows:[112,115,116]

$$\left\langle \hat{G} \right\rangle = \sum_k^P \tilde{G}(\Omega_k) + \sum_k^P \sum_{l>k}^P \tilde{G}(\Omega_k, \Omega_l),$$

$$\tilde{G}(\Omega_k) = \sum_n^s \left[ G_n(\Omega_k) + \sum_{m>n}^s G_{nm}(\Omega_k) \right], \quad \tilde{G}(\Omega_k, \Omega_l) = \sum_n^s \left[ G_n(\Omega_k, \Omega_l) + \sum_{m>n}^s G_{nm}(\Omega_k, \Omega_l) \right]$$

$$\tag{4}$$

Wherein:

$$G_n(\Omega_k) = (N_n(N_n-1)/2) \int d\vec{x}_{1,1}...\int_{\Omega_k} d\vec{r}_{n,1}\int d\sigma_{n,1} \int_{\Omega_k} d\vec{r}_{n,2}\int d\sigma_{n,2}...\int d\vec{x}_{s,N_s} \operatorname{Re}\left[\Psi^*_{MC}\hat{g}_n(\vec{r}_1,\vec{r}_2)\Psi_{MC}\right]$$

$$G_{nm}(\Omega_k) = N_n N_m \int d\vec{x}_{1,1}...\int_{\Omega_k} d\vec{r}_{n,1}\int d\sigma_{n,1}...\int_{\Omega_k} d\vec{r}_{m,1}\int d\sigma_{m,1}...\int d\vec{x}_{s,N_s} \operatorname{Re}\left[\Psi^*_{MC}\hat{g}_{nm}(\vec{r}_{n,1},\vec{r}_{m,1})\Psi_{MC}\right],$$

$$G_n(\Omega_k,\Omega_l) = (N_n(N_n-1)/2) \int d\vec{x}_{1,1}...\int_{\Omega_k} d\vec{r}_{n,1}\int d\sigma_{n,1} \int_{\Omega_l} d\vec{r}_{n,2}\int d\sigma_{n,2}...\int d\vec{x}_{s,N_s} \operatorname{Re}\left[\Psi^*_{MC}\hat{g}_n(\vec{r}_1,\vec{r}_2)\Psi_{MC}\right],$$

$$G_{nm}(\Omega_k,\Omega_l) = N_n N_m \int d\vec{x}_{1,1}...\int_{\Omega_k} d\vec{r}_{n,1}\int d\sigma_{n,1}...\int_{\Omega_l} d\vec{r}_{m,1}\int d\sigma_{m,1}...\int d\vec{x}_{s,N_s} \operatorname{Re}\left[\Psi^*_{MC}\hat{g}_{nm}(\vec{r}_{n,1},\vec{r}_{m,1})\Psi_{MC}\right].$$

The prime application of this scheme is the extension of the interacting quantum atoms (IQA) energy partitioning scheme,[118–122] to the MC systems through the partitioning of the inter-particle Coulomb interactions (vide infra).

The openness of the atomic basins with respect to the electron exchange is quantified within the context of the MC-QTAIM by introducing the basin particle number distributions, $P^n_m(\Omega_k)$, and then computing their mean, intra-basin variance and inter-basin covariance as follows:[106,116,117]

$$P^n_m(\Omega_k) = \binom{N_n}{m}\int_{\Omega_k} d\vec{r}_{n,1}...\int_{\Omega_k} d\vec{r}_{n,m} \int_{R^3-\Omega_k} d\vec{r}_{n,m+1}... \int_{R^3-\Omega_k} d\vec{r}_{n,N_n} \prod_i^{N_n}\int d\sigma_{n,i}\int d\vec{x}_{1,1}...\int d\vec{x}_{n-1,N_{n-1}}\int d\vec{x}_{n+1,1}...\int d\vec{x}_{s,N_s} \Psi^*_{MC}\Psi_{MC}$$

$$N_n(\Omega_k) = \sum_m^{N_n} m P^n_m(\Omega_k) = \int_{\Omega_k} d\vec{r}\, \rho_n(\vec{r}),$$

$$Cov_n(\Omega_k,\Omega_l) = \int_{\Omega_k} d\vec{r}_1 \int_{\Omega_l} d\vec{r}_2 \rho^{(2)}_n(\vec{r}_1,\vec{r}_2) - N_n(\Omega_k)N_n(\Omega_l)$$

$$Var_n(\Omega_k) = \sum_m^{N_n}(m-N(\Omega_k))^2 P^n_m(\Omega_k) = \int_{\Omega_k} d\vec{r}_1 \int_{\Omega_k} d\vec{r}_2 \rho^{(2)}_n(\vec{r}_1,\vec{r}_2) + N_n(\Omega_k) - [N_n(\Omega_k)]^2$$

(5)

Wherein,

$$\rho_n^{(2)}(\vec{r}_{n,1}, \vec{r}_{n,2}) = N_n(N_n - 1) \int d\vec{x}_{1,1} \ldots \int d\vec{x}_{n-1,N_n} \int d\sigma_{n,1} \int d\sigma_{n,2} \int d\vec{x}_{n,3} \ldots \int d\vec{x}_{s,N_s} \Psi_{MC}^* \Psi_{MC}$$

is the spinless reduced second-order density matrix for the $n-th$ type of particles.[106,116,117] The basin particle number distribution is the probability distribution of observing $m$-particles from the $n-th$ type within the $k-th$ basin while the rest of $N_n - m$ particles are in the rest of the basins. The index of particle delocalization is introduced as follows: $\delta_n(\Omega_k, \Omega_l) = 2|Cov_n(\Omega_k, \Omega_l)|$, which is a gauge of the openness of an atomic basin with respect to the exchange of $n-th$ type of particles with other basins.

The extended IQA method within the context of the MC-QTAIM was first developed and applied for the TC systems containing various isotopes of hydrogen as well as positronic systems using ab initio TC Hartree-Fock (TC-HF) wavefunctions.[112,115] The details of the intra-basin terms may be found in the original references,[112,115] but beyond the TC-HF approximation, for a fully correlated TC wavefunction the inter-basin terms of the extended IQA partitioning for a two-component (TC) system, composed of electrons and a proton, are as follows:

$$V_{elec}^{cl}(\Omega_k, \Omega_l) = V_{en}(\Omega_k, \Omega_l) + V_{en}(\Omega_l, \Omega_k) + V_{ee}^{cl}(\Omega_k, \Omega_l) + V_{nn}(\Omega_k, \Omega_l)$$

$$V_p^{cl}(\Omega_k, \Omega_l) = V_{ep}^{cl}(\Omega_k, \Omega_l) + V_{ep}^{cl}(\Omega_l, \Omega_k) + V_{pn}(\Omega_k, \Omega_l) + V_{pn}(\Omega_l, \Omega_k)$$

$$E_{inter}^{elec}(\Omega_k, \Omega_l) = V_{elec}^{cl}(\Omega_k, \Omega_l) + V_{elec}^{xc}(\Omega_k, \Omega_l)$$

$$E_{inter}^{total}(\Omega_k, \Omega_l) = E_{inter}^{elec}(\Omega_k, \Omega_l) + V_p^{cl}(\Omega_k, \Omega_l) + V_{ep}^c(\Omega_k, \Omega_l) + V_{ep}^c(\Omega_l, \Omega_k)$$

(6)

The details of each term are as follows:

$$V_{en}(\Omega_k,\Omega_l) = -Z_k \int_{\Omega_l} d\vec{r}_{1,1} \frac{\rho_1(\vec{r}_{1,1})}{|\vec{r}_{1,1}-\vec{R}_k|}, \quad V_{en}(\Omega_l,\Omega_k) = -Z_l \int_{\Omega_k} d\vec{r}_{1,1} \frac{\rho_1(\vec{r}_{1,1})}{|\vec{r}_{1,1}-\vec{R}_l|},$$

$$V_{ee}^{cl}(\Omega_k,\Omega_l) = \int_{\Omega_k} d\vec{r}_{1,1} \int_{\Omega_l} d\vec{r}_{1,2} \frac{\rho_1(\vec{r}_{1,1})\rho_1(\vec{r}_{1,2})}{|\vec{r}_{1,1}-\vec{r}_{1,2}|}, \quad V_{elec}^{xc}(\Omega_k,\Omega_l) = \int_{\Omega_k} d\vec{r}_{1,1} \int_{\Omega_l} d\vec{r}_{1,2} \frac{\rho_1^{xc}(\vec{r}_{1,1},\vec{r}_{1,2})}{|\vec{r}_{1,1}-\vec{r}_{1,2}|}$$

$$V_{nn}(\Omega_k,\Omega_l) = \frac{Z_k Z_l}{|\vec{R}_k-\vec{R}_l|}, \quad V_{ep}^{cl}(\Omega_k,\Omega_l) = -\int_{\Omega_k} d\vec{r}_{1,1} \int_{\Omega_l} d\vec{r}_{2,1} \frac{\rho_1(\vec{r}_{1,1})\rho_2(\vec{r}_{2,1})}{|\vec{r}_{1,1}-\vec{r}_{2,1}|}$$

$$V_{ep}^{cl}(\Omega_l,\Omega_k) = -\int_{\Omega_l} d\vec{r}_{1,1} \int_{\Omega_k} d\vec{r}_{2,1} \frac{\rho_1(\vec{r}_{1,1})\rho_2(\vec{r}_{2,1})}{|\vec{r}_{1,1}-\vec{r}_{2,1}|}, \quad , \quad V_{pn}(\Omega_k,\Omega_l) = Z_k \int_{\Omega_l} d\vec{r}_{2,1} \frac{\rho_2(\vec{r}_{2,1})}{|\vec{r}_{2,1}-\vec{R}_k|}$$

$$, V_{pn}(\Omega_l,\Omega_k) = Z_l \int_{\Omega_i} d\vec{r}_{2,1} \frac{\rho_2(\vec{r}_{2,1})}{|\vec{r}_{2,1}-\vec{R}_l|}, \quad V_{ep}^{c}(\Omega_k,\Omega_l) = -\int_{\Omega_l} d\vec{r}_{1,1} \int_{\Omega_k} d\vec{r}_{2,1} \frac{\rho_{12}^{c}(\vec{r}_{1,1},\vec{r}_{2,1})}{|\vec{r}_{1,1}-\vec{r}_{2,1}|},$$

$$V_{ep}^{c}(\Omega_l,\Omega_k) = -\int_{\Omega_k} d\vec{r}_{1,1} \int_{\Omega_l} d\vec{r}_{2,1} \frac{\rho_{12}^{c}(\vec{r}_{1,1},\vec{r}_{2,1})}{|\vec{r}_{1,1}-\vec{r}_{2,1}|} \tag{7}$$

In this partitioning scheme the spinless reduced second-order density matrix for electrons and the electron-proton pair density are decomposed as follows:

$$\rho_1^{(2)}(\vec{r}_{1,1},\vec{r}_{1,2}) = \rho_1(\vec{r}_{1,1})\rho_1(\vec{r}_{1,2}) + \rho_1^{xc}(\vec{r}_{1,1},\vec{r}_{1,2}), \quad \rho_{12}^{(2)}(\vec{r}_{1,1},\vec{r}_{2,1}) = \rho_1(\vec{r}_{1,1})\rho_2(\vec{r}_{2,1}) + \rho_{12}^{c}(\vec{r}_{1,1},\vec{r}_{2,1}),$$

respectively, where the pair density is defined as follows: $\rho_{12}^{(2)}(\vec{r}_{1,1},\vec{r}_{2,1}) = N_1 N_2 \int d\sigma_{1,1} \int d\vec{x}_{1,2}...\int d\vec{x}_{1,N_1} \int d\sigma_{2,1} \int d\vec{x}_{2,2}....\int d\vec{x}_{2,N_2} \Psi_{TC}^* \Psi_{TC}$. Let us stress, as reviewed comprehensively in a recent paper,[116] that the MC-QTAIM analysis can be performed on a large number of many-body quantum systems, containing both positively and negatively charged quantum particles, with varying masses, even when the interactions between particles are not solely of Coulombic nature.

### 2.2. Computational details

The computational scheme used for ab initio calculations in the present study is the Nuclear-Electronic Orbital (NEO) methodology developed by Hammes-Schiffer and coworkers in the last two decades.[85,89,90] The NEO includes a number of ab initio methods developed within a hierarchical framework,[89,90] similar to the usual hierarchical structure of the ab initio electronic structure theory,[5,6] which tries to solve the MC Schrödinger equation, eq. (1). The method used in the present study is the NEO density functional theory (NEO-DFT) which is basically a TC formulation of the Kohn-Sham (KS) equations of DFT.[62,123–133] The TC-KS wavefunction is a product of the electronic Slater determinant and the protonic KS spin-orbital. The computational implementation of the NEO-DFT includes the SCF solution of the TC-KS equations using the KS potentials derived from the electron-electron and electron-proton functionals.[130,131,133] All nuclei except the target proton, between the two oxygen atoms, are treated as clamped point charges.

In principle, if the proper electron-electron exchange-correlation and the electron-proton correlation functionals are employed, through simultaneous optimization of the geometry of clamped nuclei and the one-electron and the one-proton densities, the superposed states of malonaldehyde must be recovered. However, many studies revealed that this is not an achievable task with the currently available electron-proton correlation functionals.[85,91,92] Instead, full optimization through the TC-KS equations yields "broken symmetry" KS wavefunctions that are basically describing structures **1** and **3**. Even more, in the whole toolkit of the ab initio MC quantum chemistry, only very accurate MC-full-configuration-interaction derived wavefunctions, or very high-quality MC-multi-configurational SCF wavefunctions, may recover the superposed states via full optimization.[91] Such methodologies are beyond the current computational capacity for systems with a size around malonaldehyde. Inevitably, some computational tricks like the "geometry/state-averaging" must be used to first reach a symmetric geometry which would subsequently, be employed to derive the desired "delocalized" KS wavefunctions, as justified in some previous studies.[58,94] Accordingly, in the first step, a geometry for clamped nuclei in **4** and **5** was generated by the geometry-averaging method of the internal coordinates of the equilibrium geometries of **1** and **3**. The $C_s$ and $C_{2v}$ geometries of malonaldehyde, previously obtained by Hargis et al. at the CCSD(T)/cc-pVQZ level,[32] were used for the geometry averaging procedure (see the Supporting information (SI) for details). In the averaging procedure, the arithmetic mean of the corresponding Z-matrix coordinates of **1** and **3** was computed after ordering the coordinates of the two Z-matrices. Let us stress the geometry averaging procedure is not unique however, the results of the MC-QTAIM analysis, as will be discussed subsequently, are not sensitive to the details of geometry as far as the geometrical variations relative to the one used in this study is not large.

To obtain the ground and first excited one-proton wavefunctions, i.e. structures **4** and **5**, respectively, the three-dimensional time-independent Schrödinger equation of the quantum proton was solved by the Numerov method in a cubic grid of 1.75 Å edge width, using 48 grid points per dimension; the used proton mass is 1836.153 in atomic units.[61] The center of the cubic grid was placed at the point (x,y,z) = (0.0, 0.0, 0.2) in the coordinate system used to represent the averaged structures, as given in the SI. The cubic grid spanned from -0.875 Å to 0.875 Å in x and y directions and from -0.675 Å to 1.075 Å in z direction covering the most critical spatial domain for the representation of the quantum proton's wavefunction. The corresponding electronic ground-state PES for the Numerov method was derived at MP2/cc-pVTZ level at each grid point assuming a clamped proton. The high efficiency and accuracy of the *improved* Numerov method in solving high dimensional differential equations have been recently demonstrated in considering the molecular vibrational modes.[134–136] In the present study, we used the 7-point stencil that enabled us to reach the accuracy O ($10^{-7}$) in the solution of the grid-based Numerov methodology. In addition, the JADAMILU eigenvalue problem library was employed for the diagonalization procedure that takes advantage of the highly-sparse representation of the Hamiltonian formed within the Numerov approach.[137] In the next step, each numerical grid of the proton's wavefunction was fitted to a set of 6s6p6d Gaussian functions, by full optimization of their coefficients, exponents and centers (see the SI for details). The fitted proton's wavefunction was then used as an initial guess for the protonic orbital in the SCF procedure of the NEO-DFT. The cyclic updates of the protonic density were avoided during the SCF procedure to obtain the optimized electronic part of the TC-KS wavefunction at B3LYP/[cc-pVTZ:6s6p6d] level.[138,139] Accordingly, this approach could be seen as a frozen nuclear density approximation where the one-proton density remained unchanged during the optimization of the one-electron density.

Instead of the usual single-center expansion of the electronic basis functions, two sets of hydrogen cc-pVTZ basis set,[138] were placed at the two positions corresponding to the maxima of the one-proton densities (See Figure 3). Let us stress at this stage that the mentioned computational procedure has not been designed for accurately reproducing the energy gap between the ground and excited superposed protonic stats, but to infer proper TC-KS wavefunctions for the MC-QTAIM analysis. Also, in contrast to the product nature of the KS wavefunction, the fact that the electronic part of the KS wavefunction is derived self-consistently under the direct influence of the one-proton density makes the whole procedure distinct from any adiabatic procedure that treats electrons' and proton's dynamics separately.

At the next stage, and for comparison purposes, the TC-KS wavefunctions for **1** to **3**, were obtained at the two-component B3LYP:EPC17-1/cc-pVTZ:10s10p10d level,[130] by employing the CCSD(T)/cc-pVQZ optimized geometries.[32] The protonic KS spatial orbital in each of the structures **1** to **3** was expanded by [10s10p10d] uncontracted shells of even-tempered Gaussian basis functions, which are placed at the optimized position of the transferring proton and their exponents spanned the range of $2\sqrt{2}$ to $64$. In addition, the cc-pVTZ basis set was used to expand the electronic KS orbitals employing the clamped proton position derived at the CCSD(T)/cc-pVQZ level as the center of the electronic basis set. The used even-tempered protonic basis set is flexible enough to capture the anharmonicity and anisotropy of the protonic orbital, as has been shown in several recent NEO-DFT studies.[131,140] The recently developed EPC17-1 electron-proton correlation functional was used to produce high-quality one-proton density which is comparable to that derived from highly accurate ab-initio results of the MC coupled-cluster theory.[130,141] The NEO methodology was originally implemented into the GAMESS quantum chemistry package although the computations reported in the this paper have been derived via our in-house version of the NEO-GAMESS package.[85,123,142,143] Some details of the ab initio results are gathered in Table S1 in the SI.

The extended IQA partitioning of the electronic exchange-correlation energy of **1** to **5** was done according to the formalism discussed in the previous subsection and based on the IQA-DFT partitioning developed previously.[144,145] Since no electron-proton correlation functional is used in the NEO-DFT calculations of **4** and **5**, the electron-proton correlation energy is zero and the total inter-basin interaction energy reduces to: $E_{\text{inter}}^{total}(\Omega_k, \Omega_l) = E_{\text{inter}}^{elec}(\Omega_k, \Omega_l) + V_p^{cl}(\Omega_k, \Omega_l)$. The AIMALL package, T. A. Keith, AIMAll (Version 13.11.04), TK Gristmill Software, Overland Park KS, USA, 2013, was used for the MC-QTAIM calculations including the graphics of the molecular graphs and the interatomic surfaces. All computational results are offered in the atomic units both in the main text and the SI.

**3. Results and discussion**

To start the MC-QTAIM analysis, the computed TC-KS wavefunctions were used to derive the one-electron, $\rho_1(\vec{r})$, and the one-proton, $\rho_2(\vec{r})$, densities as well as various energy partition terms of the extended IQA method. Figure 3 depicts the ab initio computed $\rho_2(\vec{r})$ revealing the expected "localized" nature of the one-proton density in **1** to **3** which is in marked contrast to its "delocalized" representation in **4** and **5**. At the next stage, the MC-QTAIM partitioning was performed by employing the Gamma density, $\Gamma^{(2)}(\vec{r}) = \rho_1(\vec{r}) + (1/m_{proton})\rho_2(\vec{r})$, and the results have been summarized in Figures 4 and 5. The topological analysis of $\vec{\nabla}\Gamma^{(2)}(\vec{r})$ in **1** to **3** yields the same number and types of critical points (CPs), and thus, provides the same types of AIM that were also previously deduced from the QTAIM analysis of the ab initio electronic wavefunction of **C$_S$** and **C$_{2v}$** structures.[146] In contrast, for the superposed states, **4** and **5**, the AIM structures are quite distinct and no hydrogen basin emerges between the two oxygen basins. It seems upon delocalization in two distinct spatial positions, the proton loses its capability to attract enough electrons and form its own maxima, i.e. (3, -3) CP in $\Gamma^{(2)}(\vec{r})$, and somehow "dissolves" in the two basins containing the clamped oxygen nuclei (for the nomenclature used to classify CPs see Bader's monograph).[95] It seems that these exotic basins are better to be called the "hybrid oxygen-hydrogen" basins, hereafter denoted as *O4H* and *O5H*, to make a distinction from the usual oxygen basins in **1** to **3**. In the previous studies, this phenomenon was observed only for the light positively-charged quantum particles, e.g. positrons.[99,115]

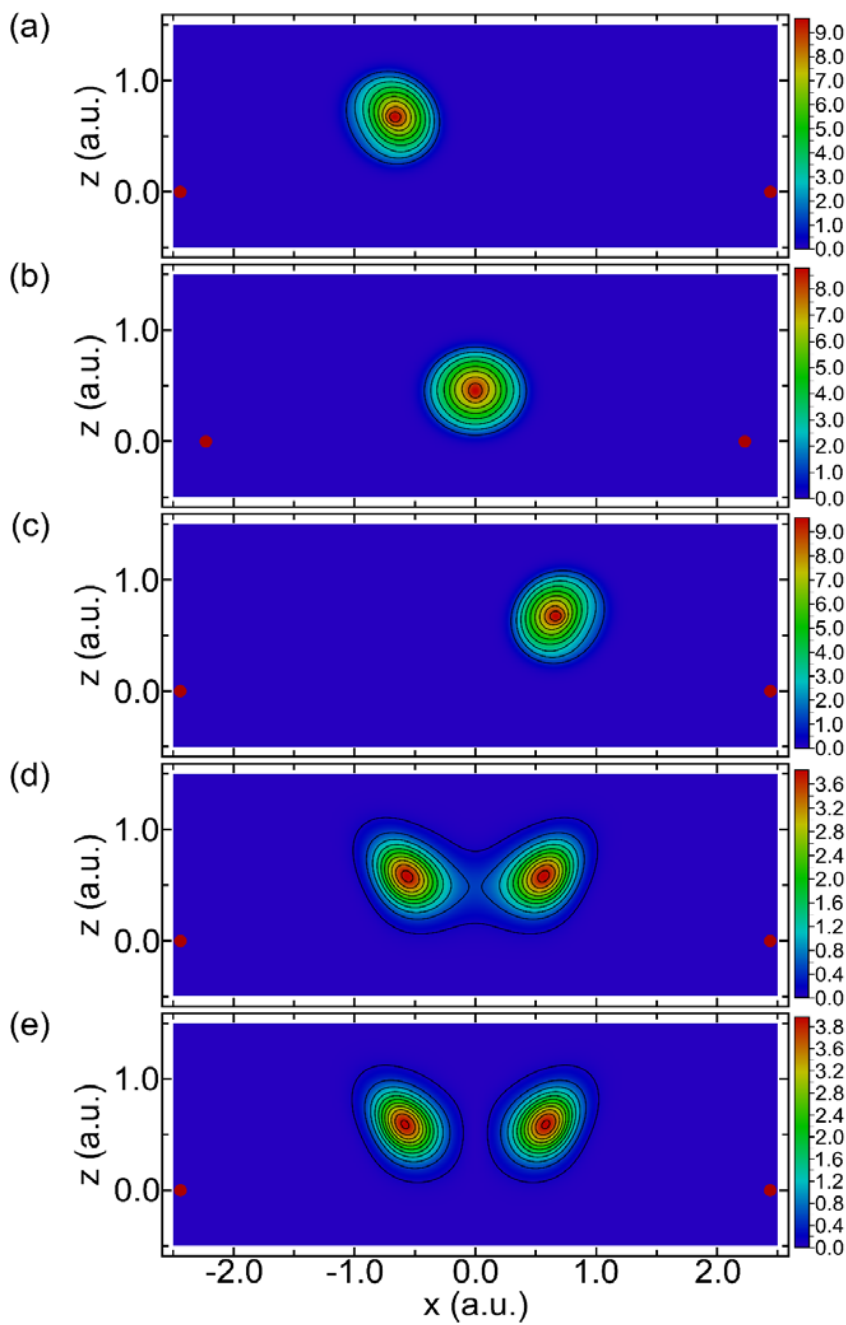

**Figure 3** 2D contour maps of the one-proton densities of (a) **1**, (b) **2**, (c) **3**, (d) **4** and (e) **5** structures. The positions of the clamped oxygen nuclei are depicted by red dots while the other nuclei are placed at the negative region of the z-axis on the xz plane (see Figure 2).

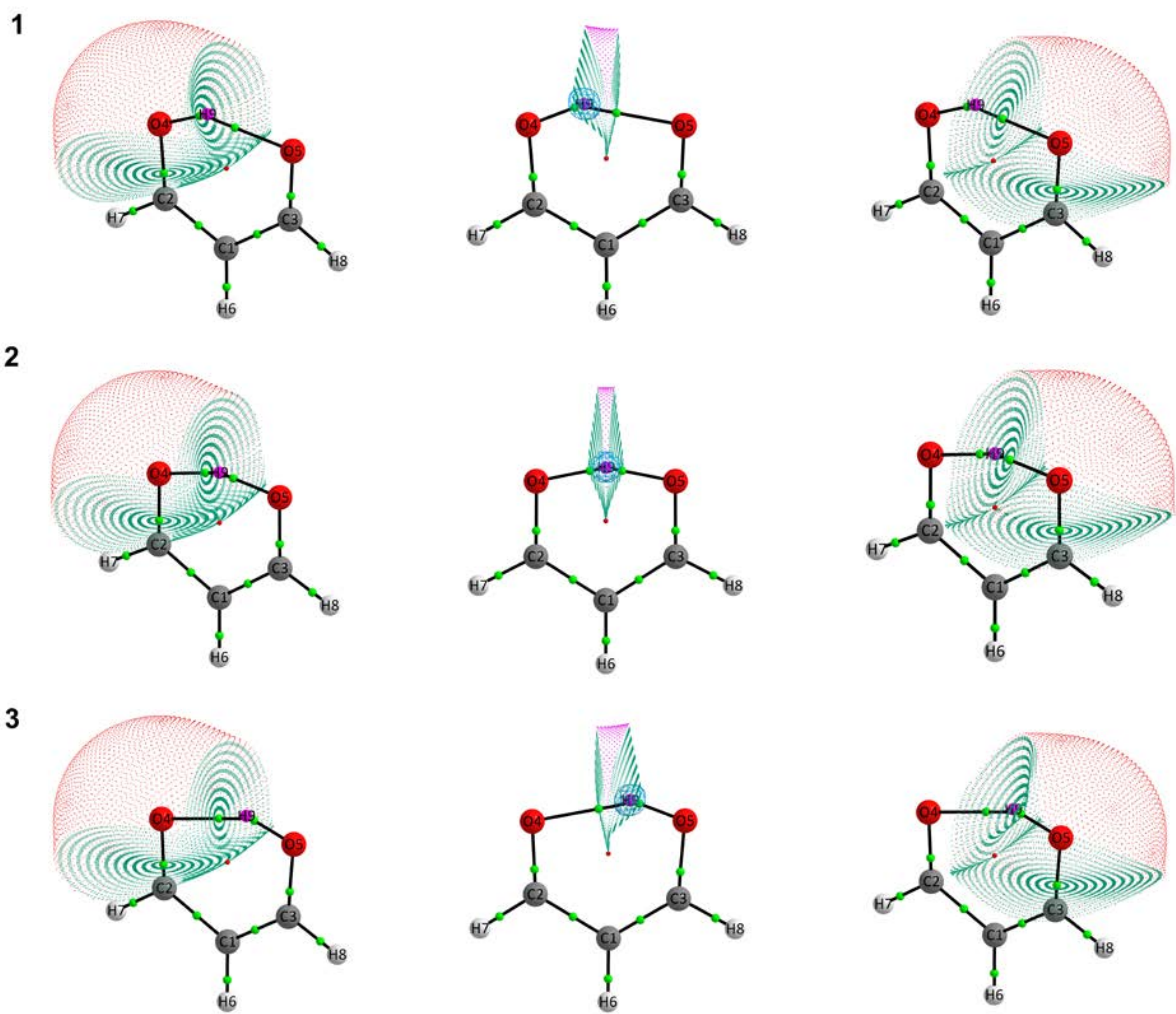

**Figure 4** The AIM structures of **1** to **3**. The blue spherical mesh is the 3D iso-density surface of the proton one-density shown at 0.0001 a.u. (for a complementary view see Figure 3) while the purple, red, and green spheres are the (3, -3), (3, +1), and (3, -1) types CPs of $\Gamma^{(2)}(\vec{r})$, respectively. The black lines are the line paths and the green surfaces are the boundaries of oxygen and hydrogen basins. The red and purple meshes show surfaces where $\Gamma^{(2)}(\vec{r})$ equals 0.001 a.u. which are an arbitrary but reasonable outer boundaries for the oxygen and hydrogen basins, respectively. The (3, -3) nuclear attractors at (or very near to) the clamped nuclei are shown by the larger spheres and labeled according to their elemental constituents.

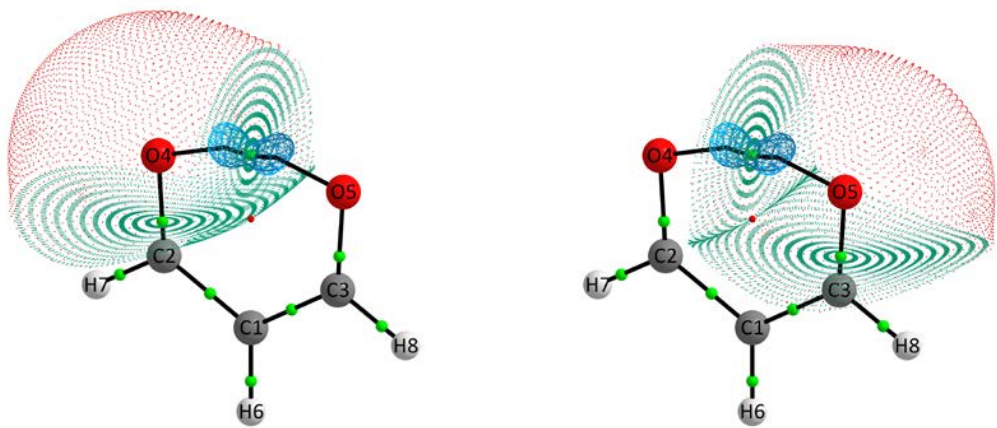

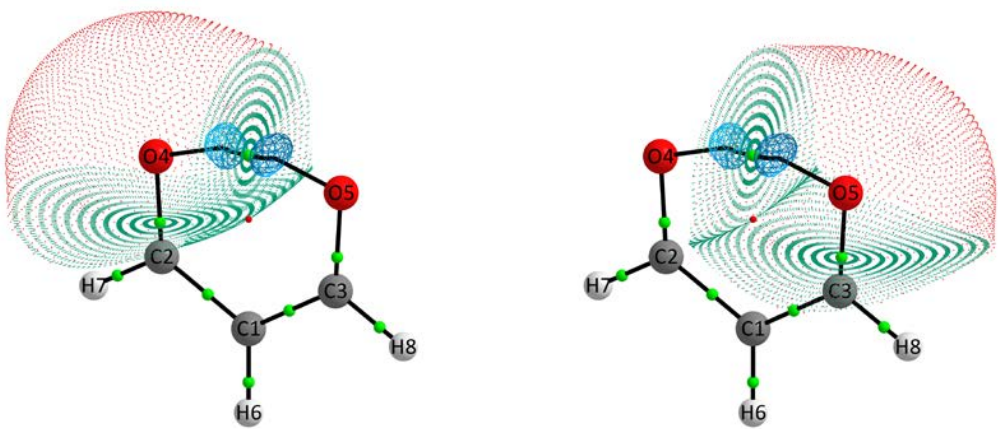

**Figure 5** The AIM structures of **4** and **5**. The blue spherical mesh in each structure is the 3D iso-density surface of the proton one-density shown at 0.0001 a.u. (for a complementary view see Figure 3) while the red and green spheres are the (3, +1), and (3, -1) types CPs of $\Gamma^{(2)}(\vec{r})$, respectively. The black lines are the line paths and the green surfaces are the boundaries of the hybrid oxygen-hydrogen basins. The red mesh shows surfaces where $\Gamma^{(2)}(\vec{r})$ equals 0.001 a.u. which are an arbitrary but reasonable outer boundaries for the hybrid oxygen-hydrogen basins. The (3, -3) nuclear attractors at (or very near to) the clamped nuclei are shown by the larger spheres and labeled according to their elemental constituents.

Table 1 contains the atomic charges, $Q(A) = Z_\Omega - N_1(A) + N_2(A)$, of all basins and the electronic delocalization indices, $\delta_1(A,B)$,[106] between the neighboring basins, i.e. those sharing an inter-atomic boundary and a line critical point,[147] in **1** to **5**. The latter index is chemically interpretable as a measure of the covalent bond order.[148] The numerical values of $\delta_1(A,B)$ imply the expected covalent networks of bonds in **1** to **3**, which are practically indistinguishable from those depicted schematically in Figure 1. On the other hand, they reveal that the molecular structures of **4** and **5** are virtually the same, and the AIM structures and bonding networks of the backbones, i.e. all the basins except the two hybrid basins, as well as the bonding between the hybrid and neighboring carbon basins, are quite similar to that of **2**. So, what remains unclear at this stage is the nature of the interaction between the hybrid basins. To dig into the nature of hybrid basins' interactions, the extended IQA energy partitioning, was applied to **1** to **5**, and the computed two-particle inter-basin terms are considered herein. The results of the extended IQA partitioning of **1** to **3**, gathered in Table 2, conform fully to the nature of bonds derived from the values of $\delta_1(A,B)$ and they may be used as the reference to deduce the nature of bonds in **4** and **5**. Table 3 offers the results of the extended IQA analysis for **4** and **5** revealing the fact that they share quite similar bonding networks in the backbone. All the bonded basin pairs, except for the hybrid pair, have stabilizing inter-basin electronic interactions, $E_{\text{inter}}^{elec}(A,B) = V_{elec}^{cl}(A,B) + V_{elec}^{xc}(A,B)$, since the inter-basin electronic exchange-correlation interaction energies, $V_{elec}^{xc}(A,B)$, are always stabilizing. Adding the total inter-basin classical Coulomb interaction of electrons and clamped nuclei with the proton does not change this picture and the total inter-basin interactions, $E_{\text{inter}}^{total}(A,B) = E_{\text{inter}}^{elec}(A,B) + V_p^{cl}(A,B)$, are also always stabilizing.

**Table 1** The atomic charges and electronic delocalization index ($\delta_1(A,B)$) of species **1** to **5** (see Figure 4 for the numbering of the atomic basins).[*]

| Atomic basin | 3 (=1) charge | 2 charge | Atomic basin | 3 (=1) $\delta_1(A,B)$ | 2 $\delta_1(A,B)$ |
|---|---|---|---|---|---|
| **C1** | -0.02 | -0.04 | | | |
| **C2** | 0.94 | 0.78 | **C1** | 1.14 | 1.32 |
| **C3** | 0.61 | 0.78 | **C1** | 1.51 | 1.32 |
| **O4** | -1.14 | -1.15 | **C2** | 1.34 | 1.19 |
| **O5** | -1.20 | -1.15 | **C3** | 1.05 | 1.19 |
| **H6** | 0.03 | 0.04 | **C1** | 0.96 | 0.97 |
| **H7** | 0.02 | 0.05 | **C2** | 0.90 | 0.91 |
| **H8** | 0.06 | 0.05 | **C3** | 0.92 | 0.91 |
| **H9** | 0.69 | 0.67 | **O5/O4** | 0.40/0.11 | 0.28 |
| | **4** | **5** | | **4** | **5** |
| **C1** | -0.05 | -0.05 | | | |
| **C2** | 0.78 | 0.78 | **C1** | 1.32 | 1.32 |
| **C3** | 0.78 | 0.78 | **C1** | 1.32 | 1.32 |
| **H6** | 0.03 | 0.03 | **C1** | 0.97 | 0.97 |
| **H7** | 0.04 | 0.04 | **C2** | 0.91 | 0.91 |
| **H8** | 0.04 | 0.04 | **C3** | 0.91 | 0.91 |
| **O4H** | -0.82 | -0.82 | **C2** | 1.20 | 1.20 |
| **O5H** | -0.82 | -0.82 | **O4H** | 0.29 | 0.28 |

**Table 2** The selected results of the extended IQA energy partitioning for **1** to **3** (see Figure 4 for the numbering of the atomic basins).

| Atomic basin | Atomic basin | 3 (=1) | | |
|---|---|---|---|---|
| | | $V_{elec}^{xc}$ | $V_{elec}^{cl}$ | $E_{inter}^{elec}$ |
| **C1** | **C3** | -0.446 | 0.064 | -0.382 |
| **C2** | **C1** | -0.352 | 0.049 | -0.304 |
| **C3** | **O5** | -0.316 | -0.510 | -0.826 |
| **H6** | **C1** | -0.285 | 0.038 | -0.247 |
| **H7** | **C2** | -0.269 | 0.059 | -0.211 |
| **H8** | **C3** | -0.277 | 0.061 | -0.216 |
| **O4** | **C2** | -0.396 | -0.722 | -1.118 |
| **H9** | **O5** | -0.127 | 0.159 | 0.032 |
| **H9** | **O4** | -0.027 | 0.111 | 0.084 |
| | | **2** | | |
| **C1** | **C2/C3** | -0.399 | 0.056 | -0.343 |
| **C3** | **O5** | -0.356 | -0.612 | -0.968 |
| **H6** | **C1** | -0.287 | 0.038 | -0.249 |
| **H7** | **C2** | -0.274 | 0.063 | -0.211 |
| **H8** | **C3** | -0.274 | 0.063 | -0.211 |
| **O4** | **C2** | -0.356 | -0.612 | -0.968 |
| **H9** | **O5/O4** | -0.084 | 0.154 | 0.069 |

**Table 3** The selected results of the extended IQA energy partitioning for **4** and **5** (see Figure 5 for the numbering of the atomic basins).

| Atomic basin | Atomic basin | $V_{elec}^{xc}$ | $V_{elec}^{cl}$ | $E_{inter}^{elec}$ | $V_p^{cl}$ | $E_{inter}^{total}$ |
|---|---|---|---|---|---|---|
| **4** | | | | | | |
| C1 | C3 | -0.398 | 0.053 | -0.345 | 0.000 | -0.345 |
| C2 | C1 | -0.398 | 0.053 | -0.345 | 0.000 | -0.345 |
| C3 | O5 | -0.358 | -0.655 | -1.013 | 0.145 | -0.868 |
| H6 | C1 | -0.286 | 0.037 | -0.249 | 0.000 | -0.249 |
| H7 | C2 | -0.274 | 0.062 | -0.212 | 0.000 | -0.212 |
| H8 | C3 | -0.274 | 0.062 | -0.212 | 0.000 | -0.212 |
| O4H | C2 | -0.358 | -0.654 | -1.012 | 0.145 | -0.867 |
| O5H | O4H | -0.055 | 0.425 | 0.370 | -0.508 | -0.138 |
| **5** | | | | | | |
| C1 | C3 | -0.398 | 0.053 | -0.345 | 0.000 | -0.345 |
| C2 | C1 | -0.398 | 0.053 | -0.345 | 0.000 | -0.345 |
| C3 | O5 | -0.358 | -0.655 | -1.013 | 0.147 | -0.866 |
| H6 | C1 | -0.286 | 0.037 | -0.249 | 0.000 | -0.249 |
| H7 | C2 | -0.274 | 0.062 | -0.212 | 0.000 | -0.212 |
| H8 | C3 | -0.274 | 0.062 | -0.212 | 0.000 | -0.212 |
| O4H | C2 | -0.358 | -0.654 | -1.012 | 0.147 | -0.866 |
| O5H | O4H | -0.053 | 0.423 | 0.370 | -0.495 | -0.124 |

In the case of the hybrid pair the inter-basin electronic exchange-correlation energy, $V_{elec}^{xc}(\Omega_{O4H},\Omega_{O5H})$, slightly stabilizing but its absolute value is very small and negligible. Whereas, the inter-basin electronic classical Coulomb interaction energy, $V_{elec}^{cl}(\Omega_{O4H},\Omega_{O5H})$, is largely destabilizing thus, the sole source of stabilizing interaction is $V_{p}^{cl}(\Omega_{O4H},\Omega_{O5H})$. The latter term itself is composed of the destabilizing interaction between the proton and the oxygen nuclei, $V_{pn}(\Omega_{O4H},\Omega_{O5H})+V_{pn}(\Omega_{O5H},\Omega_{O4H})$, and the stabilizing classical Coulomb interaction energy of the one-proton density in one of the basins with the one-electron density in the other basin, $V_{ep}^{cl}(\Omega_{O4H},\Omega_{O5H})+V_{ep}^{cl}(\Omega_{O5H},\Omega_{O4H})$. Thus, the very origin of the stabilizing interaction between the hybrid basins is the latter terms in $V_{p}^{cl}(\Omega_{O4H},\Omega_{O5H})$, and it seems reasonable to claim that these terms yield an unconventional chemical bond between the two hybrid basins. Let us stress that this mechanism of bonding is unprecedented in usual molecules but it is interestingly quite similar to the mechanism of formation of the newly proposed "positronic bond".[115,149]

## 4. Conclusion

Malonaldehyde is usually conceived as a typical organic molecule with a molecular structure consisted of a network of well-known C-C, C-O, C-H and O-H covalent bonds, and an intra-molecular resonant-assisted hydrogen bond.[150] Even susceptibility to the tunneling mechanism does not change this picture profoundly and adds an extra dynamical, time-dependent, aspect to its molecular structure.[76] However as demonstrated in the present study, applying the quantum superposition principle to the proton between the oxygen atoms alters significantly its molecular structure. In fact, the very basic and probably trivial idea that each atom is a node in the molecular graph is called into question in the quantum superposed states, at least within the context of the present MC-QTAIM analysis. Evidently, a proper symbolic representation of molecular structures attributable to such exotic states is lacking in the current state of affairs and needs novel conventions and nomenclature. Probably, the fruitfulness of attributing molecular structures to such exotic states may be evaluated if the corresponding "exotic chemistry" will be amenable to future experimental studies. Indeed, the analysis of the AIM structure and the bonding network of the superposed nuclear states is a completely uncharted territory that awaits further theoretical and experimental developments and computational analysis.

**Acknowledgments**

The authors are grateful to Cina Foroutan-Nejad for his constructive comments.

# Supporting Information

## The MC-QTAIM analysis reveals an exotic bond in the coherently quantum superposed Malonaldehyde

Mohammad Goli[1] and Shant Shahbazian[2]

[1]School of Nano Science, Institute for Research in Fundamental Sciences (IPM), Tehran 19395-5531, Iran,

Email: m_goli@ipm.ir

[2]Department of Physics, Shahid Beheshti University, Evin, Tehran, Iran

E-mail: sh_shahbazian@sbu.ac.ir

## Table of contents

| The optimized geometry of **1** (or **3**) at the CCSD(T)/cc-pVQZ level ||||||
| Atom | Number | Coordinates [Bohr] |||
| | | X | Y | Z |
| C | 1 | 0.000000 | 2.077361 | 0.000000 |
| C | 2 | 2.339522 | 0.675322 | 0.000000 |
| C | 3 | -2.236791 | 0.811101 | 0.000000 |
| O | 4 | 2.439903 | -1.657362 | 0.000000 |
| O | 5 | -2.441842 | -1.681011 | 0.000000 |
| H | 6 | 0.019948 | 4.116154 | 0.000000 |
| H | 7 | 4.109976 | 1.763188 | 0.000000 |
| H | 8 | -4.041823 | 1.778489 | 0.000000 |
| H* | 9 | -0.688985 | -2.333555 | 0.000000 |

* The coordinates of the quantum hydrogen atom

| The optimized geometry of **2** at the CCSD(T)/cc-pVQZ level ||||||
| Atom | Number | Coordinates [Bohr] |||
| | | X | Y | Z |
| C | 1 | 0.000000 | 0.000000 | 2.130982 |
| C | 2 | 0.000000 | -2.239563 | 0.731188 |
| C | 3 | 0.000000 | 2.239563 | 0.731188 |
| O | 4 | 0.000000 | -2.229682 | -1.682393 |
| O | 5 | 0.000000 | 2.229682 | -1.682393 |
| H | 6 | 0.000000 | 0.000000 | 4.166644 |
| H | 7 | 0.000000 | -4.079697 | 1.658786 |
| H | 8 | 0.000000 | 4.079697 | 1.658786 |
| H* | 9 | 0.000000 | 0.000000 | -2.126078 |

* The coordinates of the quantum hydrogen atom

| The averaged geometry of **4** | | | | |
|---|---|---|---|---|
| Atom | Number | Coordinates [Bohr] | | |
| | | X | Y | Z |
| C | 1 | 0.000000 | 0.000000 | 4.458367 |
| C | 2 | 2.288822 | 0.000000 | 3.124956 |
| C | 3 | -2.288822 | 0.000000 | 3.124956 |
| O | 4 | 2.439951 | 0.000000 | 0.711995 |
| O | 5 | -2.439951 | 0.000000 | 0.711995 |
| H | 6 | 0.000000 | 0.000000 | 6.497257 |
| H | 7 | 4.077565 | 0.000000 | 4.152645 |
| H | 8 | -4.077565 | 0.000000 | 4.152645 |
| Bq* | 9 | -0.564373 | 0.000000 | 0.132820 |
| Bq* | 10 | 0.564373 | 0.000000 | 0.132820 |

* The coordinates of the "ghost atoms" used as the centers of expansion for the electronic basis functions of the quantum hydrogen atom coinciding with the maxima of the ground state protonic orbital.

| The averaged geometry of **5** | | | | |
|---|---|---|---|---|
| Atom | Number | Coordinates [Bohr] | | |
| | | X | Y | Z |
| C | 1 | 0.000000 | 0.000000 | 4.458367 |
| C | 2 | 2.288822 | 0.000000 | 3.124956 |
| C | 3 | -2.288822 | 0.000000 | 3.124956 |
| O | 4 | -2.439951 | 0.000000 | 0.711995 |
| O | 5 | 2.439951 | 0.000000 | 0.711995 |
| H | 6 | 0.000000 | 0.000000 | 6.497257 |
| H | 7 | 4.077565 | 0.000000 | 4.152645 |
| H | 8 | -4.077565 | 0.000000 | 4.152645 |
| Bq* | 9 | -0.582494 | 0.000000 | 0.121241 |
| Bq* | 10 | 0.582494 | 0.000000 | 0.121241 |

* The coordinates of the "ghost atoms" used as the centers of expansion for the electronic basis functions of the quantum hydrogen atom coinciding with the maxima of the excited state protonic orbital.

# The regression procedure: Transforming the numerical protonic wavefunctions into the Gaussian-type orbitals

The numerical ground and excited protonic orbitals, obtained by the 3D Numerov method, were fitted by linear combinations of [6s6p6d] Cartesian Gaussian-type shells which were placed on the molecular symmetry plane of the averaged geometry that contains all the clamped nuclei. This is called the xz plane where the x-axis is the $C_2$ geometrical symmetry axis with respect to the rotation of the clamped nuclei. In the fitting procedure, the protonic orbital is explicitly shown by:

$$\Psi_{ground\ state} = \sum_{i=1}^{15}\left[c_i\varphi_{\alpha_i}(x_i, y_i, z_i) + c_i\varphi_{\alpha_i}(-x_i, y_i, z_i)\right],$$

and,

$$\Psi_{excited\ state} = \sum_{i=1}^{15}\left[c_i\varphi_{\alpha_i}(x_i, y_i, z_i) - c_i\varphi_{\alpha_i}(-x_i, y_i, z_i)\right],$$

where $c_i$, $\alpha_i$ and $(x_i, y_i, z_i)$ are the coefficients, exponents and coordinates of the center of *i-th normalized* Gaussian function in these series, respectively. It must be noted that only non-zero contributions, due to the molecular symmetry, were included in the wavefunction expansions and offered in the forthcoming tables of the optimal parameters. The following tables gather the optimized parameters of the protonic orbitals, which were derived after the fitting of the Gaussian expansions to the 3D grid of the protonic wavefunctions.

| Regression parameters of the ground state protonic orbital ||||||||
| Function Number | Shell Number | Function Type | Exponent | Coefficient | Coordinates [Bohr] |||
| | | | | | X | Y | Z |
|---|---|---|---|---|---|---|---|
| 1 | 1 | S | 5.92536 | -0.667308 | 0.098133 | 0.000000 | -0.042515 |
| 2 | 2 | S | 4.90609 | 0.103662 | 0.831002 | 0.000000 | -0.082282 |
| 3 | 3 | S | 5.53872 | 0.971034 | 0.256124 | 0.000000 | -0.022740 |
| 4 | 4 | Pz | 9.99765 | -0.025588 | 0.894294 | 0.000000 | -0.102668 |
| 5 | 5 | Pz | 5.98643 | 0.277155 | 0.488342 | 0.000000 | -0.086228 |
| 6 | 6 | Pz | 8.71275 | -0.067002 | 0.668808 | 0.000000 | 0.221253 |
| 7 | 7 | Dxx | 6.17956 | 0.583140 | 0.559992 | 0.000000 | 0.043214 |
| 8 | 7 | Dyy | 6.17956 | -0.405039 | 0.559992 | 0.000000 | 0.043214 |
| 9 | 7 | Dzz | 6.17956 | 0.505316 | 0.559992 | 0.000000 | 0.043214 |
| 10 | 8 | Dxx | 6.97638 | 0.417660 | 0.736994 | 0.000000 | 0.148422 |
| 11 | 8 | Dyy | 6.97638 | -0.295682 | 0.736994 | 0.000000 | 0.148422 |
| 12 | 8 | Dzz | 6.97638 | 0.230332 | 0.736994 | 0.000000 | 0.148422 |
| 13 | 9 | Dxx | 6.33365 | -0.999577 | 0.649366 | 0.000000 | 0.093094 |
| 14 | 9 | Dyy | 6.33365 | 0.756131 | 0.649366 | 0.000000 | 0.093094 |
| 15 | 9 | Dzz | 6.33365 | -0.587118 | 0.649366 | 0.000000 | 0.093094 |

| Regression parameters of the excited state protonic orbital ||||||||
| Function Number | Shell Number | Function Type | Exponent | Coefficient | Coordinates [Bohr] |||
| | | | | | X | Y | Z |
|---|---|---|---|---|---|---|---|
| 1 | 1 | S | 7.94753 | -0.995443 | 0.063433 | 0.000000 | -0.199071 |
| 2 | 2 | S | 8.43716 | 0.469871 | 0.967696 | 0.000000 | 0.165908 |
| 3 | 3 | S | 6.71810 | 0.999654 | 0.280224 | 0.000000 | -0.116529 |
| 4 | 4 | Pz | 6.37756 | -0.778184 | 0.720576 | 0.000000 | 0.060806 |
| 5 | 5 | Pz | 7.81809 | 0.835965 | 0.545553 | 0.000000 | 0.019451 |
| 6 | 6 | Pz | 7.93854 | 0.155353 | 0.834297 | 0.000000 | 0.249593 |
| 7 | 7 | Dxx | 7.17292 | 0.272671 | 0.498309 | 0.000000 | 0.080944 |
| 8 | 7 | Dyy | 7.17292 | 0.310717 | 0.498309 | 0.000000 | 0.080944 |
| 9 | 7 | Dzz | 7.17292 | 0.434034 | 0.498309 | 0.000000 | 0.080944 |
| 10 | 8 | Dxx | 6.66305 | -0.207159 | 0.560419 | 0.000000 | 0.028372 |
| 11 | 8 | Dyy | 6.66305 | -0.999654 | 0.560419 | 0.000000 | 0.028372 |
| 12 | 8 | Dzz | 6.66305 | -0.402189 | 0.560419 | 0.000000 | 0.028372 |
| 13 | 9 | Dxx | 6.30875 | -0.468891 | 0.618625 | 0.000000 | 0.067678 |
| 14 | 9 | Dyy | 6.30875 | 0.861386 | 0.618625 | 0.000000 | 0.067678 |
| 15 | 9 | Dzz | 6.30875 | 0.019168 | 0.618625 | 0.000000 | 0.067678 |

**Table S1** - Summary of the ab initio NEO-DFT results.

| Method/basis set | B3LYP-EPC17-1/[cc-pVTZ:10s10p10d] | |
|---|---|---|
| System | 2 | 1 and 3 |
| *Electronic kinetic energy* | 265.577772 | 265.554369 |
| *Proton's kinetic energy* | 0.007718 | 0.007746 |
| *Electron-electron potential energy* | 261.422933 | 258.206955 |
| *Proton-electron potential energy* | -13.233539 | -12.654149 |
| *electron-clamped nuclei potential energy* | -949.386639 | -943.531760 |
| *Proton-clamped nuclei potential energy* | 12.255060 | 11.684860 |
| *Clamped Nuclei-clamped Nuclei potential energy* | 156.082444 | 153.453666 |
| Total energy | -267.274253 | -267.278314 |
| Virial ratio | 2.006359 | 2.006463 |
| **Method/basis set** | **B3LYP/[cc-pVTZ:6s6p6d]** | |
| System | 4 | 5 |
| *Electronic kinetic energy* | 265.443959 | 265.454513 |
| *Proton's kinetic energy* | 0.005092 | 0.005817 |
| *Electron-electron potential energy* | 258.213390 | 258.215910 |
| *Proton-electron potential energy* | -12.401579 | -12.396164 |
| *electron-clamped nuclei potential energy* | -943.513438 | -943.531044 |
| *Proton-clamped nuclei potential energy* | 11.709815 | 11.712679 |
| *Clamped Nuclei-clamped Nuclei potential energy* | 153.388670 | 153.388670 |
| *Total energy* | -267.154092 | -267.149620 |
| *Virial ratio* | 2.006423 | 2.006364 |